\documentstyle[12pt,aasms4]{article}
\newcommand{\beq}{\begin{equation}}
\newcommand{\beqa}{\begin{eqnarray}}
\newcommand{\eeq}{\end{equation}}
\newcommand{\eeqa}{\end{eqnarray}}
\newcommand{\simg}{\gtrsim}
\newcommand{\siml}{\lesssim}
\newcommand{\sol}{M_\odot}

\begin{document}
%
%{KUNS-1562,OU-TAP 94}
\title{ Density Distribution and Shape  of Galactic Dark Halo Can be Determined \\
  by Low Frequency Gravitational Waves ?}
\author{
Kunihito Ioka,$^{1}$
Takahiro Tanaka,$^{2}$
and Takashi Nakamura$^{3}$
}
\affil{$^{1}$ Department of Physics, Kyoto University, Kyoto 606-8502,
Japan}
\affil{$^{2}$Department of Earth and Space Science, Osaka 
University, Toyonaka 560-0043, Japan }
\affil{
$^{3}$Yukawa Institute for Theoretical Physics, Kyoto University, 
Kyoto 606-8502, Japan}
\centerline{Mar~~1~~1999}
\authoremail{iokakuni@tap.scphys.kyoto-u.ac.jp}

\begin{abstract}
   Under the assumption that the Milky Way's dark halo consists of
   primordial black hole MACHOs (PBHMACHOs),
   the mass density of the halo can be measured
   by the low frequency gravitational waves
   ($10^{-3}$ Hz $\siml \nu_{gw} \siml 10^{-1}$ Hz)
   from  PBHMACHO binaries whose fraction is $\sim 10^{-6}$.
   We find that ten years observation by LISA will detect
   $\sim 700$ PBHMACHO binaries and  enable us to determine the
   power index of the density profile within 10\% (20\%) 
   and the core radius within 25\% (50\%)  
   in about 90\% (99\%) confidence level, respectively.
   The axial ratios of the halo  may also be determined within $\sim 10\%$.
    LISA and OMEGA may give us 
   an unique observational method to determine
   the density profile and the shape of the dark halo to open a new
   field of observational astronomy.
\end{abstract}
 

\section{INTRODUCTION}
It is important to determine the density profile of the Milky Way's 
dark halo observationally
in order to gain insights into the galaxy formation and evolution.
Unfortunately, little has been known about the halo density profile
(HDP),
since the dark halo emits little light.
A  HI rotation curve tells us about the radial distribution of
dark matter (e.g. Fich \& Tremaine 1991). However
it has not been accurately measured for $r > D_0 \sim 8.5$ kpc and
recently there is also an argument that the Galactic rotation curve may deviate
from that of the standard halo model (Honma \& Sofue 1996, 1997,
Olling \& Merrifield 1998).
The dynamics of satellite galaxies and globular clusters
can  provide the mass inside $r\sim 50$ kpc
with some biases (e.g. Lin, Jones, \& Klemola 1995,
Kochanek 1996, Zaritsky et al. 1989, Einasto \& Lynden-Bell 1982,
Peebles 1995).
The HDP may be recovered 
by using the tidal streams from Galactic satellites 
(e.g. Johnston et al. 1999).
All methods so far are, however, indirect methods. 
%All methods so far to reconstruct the HDP,
%such as a HI rotation curve, the dynamics of satellite galaxies
%and so on, are indirect methods
%(Fich \& Tremaine 1991, Honma \& Sofue 1996, 1997,
%Olling \& Merrifield 1998, Lin, Jones, \& Klemola 1995,
%Kochanek 1996, Zaritsky et al. 1989, Einasto \& Lynden-Bell 1982,
%Peebles 1995, Johnston et al. 1999).
In this paper we investigate a possibility of direct
measurement of  the density distribution of Galactic dark halo.

Recently, Ioka, Tanaka, \& Nakamura (1998b) (hereafter ITN)
proposed a possibility to determine a map of a HDP
by low frequency gravitational waves
($10^{-4}$--$10^{-1}$ Hz) from PBHMACHO binaries,
which can be detected by the planned Laser
Interferometer Space Antenna (LISA) (Bender et al. 1998)
and Orbiting Medium Explorer for Gravitational Astrophysics (OMEGA).
ITN was motivated by
the observations of gravitational microlensing toward the Large
Magellanic Cloud (LMC).
The analysis of the first $2.1$ years of photometry of $8.5 \times 10^6$
stars in the LMC by the MACHO Collaboration (Alcock et al. 1997)
suggests that the fraction $0.62^{+0.3}_{-0.2}$
of the halo consists of massive compact halo objects (MACHOs)
of mass $0.5^{+0.3}_{-0.2} M_{\odot}$
assuming the standard spherical flat rotation halo model.
%At present, however, none of candidates,
%such as white dwarfs, LMC-LMC self-lensing and so on,
%convincingly explain the microlensing events toward LMC and SMC
%(Chabrier 1999, Gould, Flynn, \& Bahcall 1998, Bennett 1998,
%see also references in ITN,
%see also discussions on SMC events,
%Afonso et al. 1998, Albrow et al. 1998, Alcock et al. 1998,
%Honma 1999).
At present, we do not know what MACHOs are.
There are several candidates proposed to explain MACHOs,
such as brown dwarfs, red dwarfs, white dwarfs and so on
(Chabrier 1999, Gould, Flynn, \& Bahcall 1998,
see also references in ITN).
Any objects clustered somewhere between the LMC and the sun 
with the column density larger than  $25\sol{\rm pc}^{-2}$
can also explain the data (Nakamura, Kan-ya, \& Nishi 1996).
They include the possibilities:
LMC-LMC self-lensing, the thick disk, warps, tidal debris and so on
(Sahu 1994, Zhao 1998a,b, Evans et al. 1998, Gates et al. 1998,
Bennett 1998, see also discussions on SMC events,
Afonso et al. 1998, Albrow et al. 1998, Alcock et al. 1998,
Honma 1999)
However, none of them do not convincingly explain the microlensing
events toward the LMC and SMC.

Freese, Fields and Graff (1999) claimed that on theoretical grounds
one is pushed to either exotic explanations or a non-MACHO halo.
We here simply adopt the suggestion by the MACHO Collaboration
(Alcock et al. 1997) and consider an example of exotic explanations: primordial
black hole MACHO (Nakamura et al. 1997).
This possibility is free from observational constraints at present
(Fujita et al. 1998) and  PBHMACHOs may  be identified by
LIGO, VIRGO, TAMA and GEO within next 5 years
(Nakamura et al. 1997, Ioka et al. 1998a), if they exist as  dark matter.

If primordial black holes (PBHs) were formed
in the early universe at $t\sim 10^{-5}$ s
(Yokoyama 1997, Kawasaki \& Yanagida 1999, Jedamzik 1997),
a part of them evolved into binaries through the three body interactions 
(Nakamura et al. 1997, Ioka et al. 1998a).
Some of these binaries emit gravitational waves (GWs)
at low frequencies at present.
ITN found that
one year observation by LISA will be able to
identify at least several hundreds of PBHMACHO binaries.
Since LISA can measure distances and positions of PBHMACHO binaries
(Bender et al. 1998, Cutler 1998),
it may be possible to obtain  HDP
from the distribution map of PBHMACHO binaries.

In this paper, we will quantitatively
investigate how well HDP can be determined
by the observation of the low frequency GWs from PBHMACHO binaries
and show that LISA and OMEGA will serve as  excellent instruments
for determination of  the shape of our dark halo.

\section{PBHMACHO MODEL}
For simplicity, we assume that PBHs dominate the dark matter,
i.e., $\Omega=\Omega_{BH}$,
where $\Omega_{BH}$ is the density parameter of PBHs at present,
and that all PBHs have the same mass $M_{BH}$.
Throughout this paper, we will
set $M_{BH}=0.5M_{\odot}$ and $\Omega h^2=0.1$,
where $h$ is the present Hubble parameter $H_0$
in unit of $100$ km s$^{-1}$ Mpc$^{-1}$.

Assuming that PBHs are distributed randomly at their formation,
we can obtain the probability distribution function (PDF)
for orbital frequency $\nu_p$
and eccentricity 
$e$ of the binary, $f_{\nu,e}(\nu_p,e) d\nu_p de$
(Nakamura et al. 1997, Ioka et al. 1998a).
For $e \ll 1$,
an approximate PDF  is given by
\begin{equation}
 f_{\nu,t}(\nu_p;t_0)d\nu_p\sim 
 {425\over 3552} \left({{t_0}\over{\bar t}}\right)^{3/37}
 \left({{a}\over{a_0}}\right)^4
 \Gamma\left({{58}\over{37}}\right) {{d\nu_p}\over{\nu_p}},
 \label{fnuapp}
\end{equation}
where $a=(G M_{BH}/2 \pi^2 \nu_p^2)^{1/3}$ is the semimajor axis,
$t_0=10^{10}$ years is the age of the universe, 
$a_0=2.0\times 10^{11} (M_{BH}/M_{\odot})^{3/4}$ cm
is the semimajor axis of a binary in a circular orbit which coalesces
in $t_0$,
$\bar{x}=1.2\times 10^{16}(M_{BH}/M_{\odot})^{1/3}(\Omega h^2)^{-4/3}$ cm
is the mean separation of black holes at the time of matter-radiation
equality
and $\bar t=\beta^7 ({{\alpha \bar x}/{a_0}})^4 t_0$
(ITN).
$\alpha$ and $\beta$ are constants of order unity.
In this paper we adopt $\alpha=0.5$ and $\beta=0.7$
(ITN).
Note that a circular binary with orbital frequency $\nu_p$ 
emits GWs at GW frequency
$\nu_{gw}=p \nu_p$ with the second harmonic
$p=2$ (Peter \& Mathew 1963, Hils 1991).
%$f_{\nu,t}(\nu_p;t_1)$ is proportional to $\nu_p^{-11/3}$,

\section{INDIVIDUALLY OBSERVABLE SOURCES}
To be observed as individual sources,
the amplitudes of the GWs from the binaries have to
exceed the threshold amplitude $h^{th} = 5 h_{\nu} (\Delta {\nu})^{1/2}$
which is determined by the GW background $h_{\nu}$
and the frequency resolution $\Delta {\nu}=1/T$ (Schutz 1997, Thorne 1987).
Here $T$ is the observation time, and we set the 
signal-to-noise ratio ($SNR$)
as $5$.\footnote{
For simplicity we do not consider  effects of  the inclination of binaries 
and a reduction factor due to the antenna pattern of the detector in details.
These effects can be absorbed in the observation time $T$,
and the conclusion of this paper will hold 
if the effective $T$ is increased correspondingly.
}
In our model, the GW background $h_{\nu}$ is determined by
PBHMACHO binaries themselves (ITN, Hiscock 1998).
We use Figure 6 in ITN to estimate the GW background $h_{\nu}$.
The amplitude of the GW at $\nu_{gw}$ 
from a binary with eccentricity $e$, the harmonic $p$
and the distance from the earth $d$ is given by
$h_i={{2\sqrt{G F_i/c^3 \pi \nu_{gw}^2}}}$,
where $F_i={L^{(p)}(\nu_{gw},e)/{4\pi d^2}}
:={L_0 \nu_{gw}^{10/3} p^{-10/3} g(p,e)}/4 \pi d^2$ is the GW flux
and $L_0=(32 c^5/5 G)(2 \pi G {\cal M}/c^3)^{10/3}$.
%$L_0=(32 c^5/5 G)(\pi \nu_{gw} G {\cal M}/c^3)^{10/3}$.
The function $g(p,e)$ is given by Peter \& Mathew (1963),
and ${\cal M}=M_{BH}/2^{1/5}$ is the charp mass.
Then, the requirement that the signal exceeds the threshold, 
$h_i>h^{th}$, determines 
the maximum distance to individually observable sources 
with $p=2$ and $e=0$ as
\begin{equation}
%  D^{(p)}_{max}(\nu_{gw},e)
%  = {{\sqrt{G}}\over{\pi c^{3/2} h^{th}}}
%  \sqrt{L_0 g(p,e)} \nu^{2/3}_{gw} p^{-5/3}.
D^{(p)}_{max}(\nu_{gw},e)=87.2 
\left({{h^{th}}\over{10^{-23}}}\right)^{-1}
%\left({{\cal M}\over{M_{\odot}/2^{6/5}}}\right)^{5/3}
\left({{\nu_{gw}}\over{10^{-2} {\rm Hz}}}\right)^{2/3}
%\left({{p}\over{2}}\right)^{-5/3} 
%\left({{g(p,e)}\over{1}}\right)^{1/2}
{\rm kpc}.
  \label{dmax}
\end{equation}

At frequencies above $\sim \nu_{dis}=10^{-2.23}$ $(T/1{\rm
  year})^{-6/11}$ $(M_{BH}/ 0.5M_{\odot})^{-5/11}$ Hz,
LISA can measure distances to PBHMACHO binaries
since binaries change their GW frequencies $\nu_{gw}$
by more than $\Delta \nu$ through GW emission within the 
observation time $T$ (Bender et al. 1998, Schutz 1986).
%Since
%binaries with high orbital frequencies ($\nu_p \simg 10^{-3}$ Hz)
%are almost in circular orbits at present (ITN),
%we can assume that the binaries
%to which LISA can measure distances are in circular orbits
%($e=0$ and $p=2$).
Since binaries with high orbital frequencies ($\nu_p \simg 10^{-3}$ Hz)
are almost in circular orbits at present (ITN),
we can assume $e=0$ and $p=2$.
If a source with a circular orbit
changes its GW frequency by $\xi \Delta \nu = \xi/T$ 
during the observation,
the change in the binding energy of the binary is given by
$\Delta E = (c^5 \xi/3\pi \nu_{gw}^2 G T)(\pi \nu_{gw} 
G {\cal M}/c^3)^{5/3}$.
{}From the relation $L^{(2)}(\nu_{gw},0) = \Delta E / T$,
we can obtain the charp mass as 
${\cal M}=(5\pi\xi/96)^{3/5}(c^3/\pi\nu_{gw} G)(\pi \nu_{gw} T)^{-6/5}$.
Substituting the charp mass to the amplitude
$h_i=(32/5)^{1/2}(c/\pi\nu_{gw} d)(\pi\nu_{gw} G {\cal M}/c^3)^{5/3}$,
we can obtain the distance to the source,
$d=(5/288)^{1/2} (c \xi / \pi^2 \nu_{gw}^3 h_i T^2)$,
as a function of $h_i$, $\nu_{gw}$, and $\xi$.

The observable parameters, $h_i$, $\nu_{gw}$ and $\xi$,
contain observational errors as,
$h_i(1\pm\epsilon_1)$, $\nu_{gw}(1\pm\epsilon_2)$ and
$\xi(1\pm\epsilon_3)$,
which can be estimated as
$\epsilon_1=1/(SNR)$, $\epsilon_2=1/\nu_{gw}T$ and $\epsilon_3=1/\xi$.
Therefore the observational error in the distance can be estimated as
$d[1\pm(\epsilon_1+3\epsilon_2+\epsilon_3)]$.
The angular resolution of LISA is estimated to be a few degree
(Cutler 1998)\footnote{
The results of Cutler (1998) are only valid for large $SNR$
and the angular resolution may be worse in a realistic detection with
$SNR=5$ (Balasubramanian, Sathyaprakash, \& Dhurandhar 1996),
although we here simply adopt the Cutler's results.
Note also that the relative velocities of the sources
to the solar system are not taken into account in Cutler (1998).}.

\section{DENSITY PROFILE RECONSTRUCTION}
In this section  we show one of simulations of real observations.
For simplicity, we assume that the distribution of 
the number density of PBHMACHOs obeys the law  as 
\begin{equation}
  n(r)={n_s\over [1+(r^2/D_a^2)]^{\lambda}},
  \label{nr}
\end{equation}
where $r$, $n_s$, $D_a$ and $\lambda$ are the galactcentric radius,
the number density of PBHMACHOs at the galactic center,
the core radius and the power index, respectively.
As the ``real'' parameters for a simulation we set 
$n_s=2.60 \times 10^{-2}$ pc$^{-3}$, $D_a=5$ kpc and $\lambda=1$.
The total number of PBHMACHOs within $r<D_{halo}$ is given by
$N_{total}=\int_{r<D_{halo}} n(r) d^3x$ where $D_{halo}$
is the size of the halo .
Since the fraction of PBHMACHO binaries with $(10^{-3}$ Hz 
$\siml) \nu_{min}<\nu_{gw}<\nu_{max}$
is given by $F_b(\nu_{min},\nu_{max}):
=\int^{\nu_{max}}_{\nu_{min}} f_{\nu,t}(\nu_{p};t_0)
(d\nu_{p}/d\nu_{gw}) d\nu_{gw}$ with $p=2$
from equation (\ref{fnuapp}),
the total number of PBHMACHO binaries with $(10^{-3}$ Hz
$\siml) \nu_{min}<\nu_{gw}<\nu_{max}$
is given by $N_b=N_{total} F_b(\nu_{min},\nu_{max})$.
Note that, as long as $\nu_{gw} \simg 10^{-3}$ Hz,
it is sufficient to consider the case $e \ll 1$ 
and the second harmonic $p=2$.
For example, 
$N_{total}=4.03\times 10^{12}$, 
$F_b(\nu_{min},\nu_{max})=2.14 \times 10^{-8}$ and 
hence $N_{b}=8.62\times 10^4$,
for $D_{halo}=500$ kpc, $\nu_{min}=4\times 10^{-3}$ Hz, 
and $\nu_{max}=1\times 10^{-1}$ Hz.

The following algorithm explains a method of our simulations 
to see how well HDP can be determined
by the observation of the low frequency GWs.

\begin{enumerate}
%\item  We consider a HDP that depends on some parameters, and
%set parameters as the ``real'' ones.
%For example we adopt the HDP in equation (\ref{nr}).

\item We distribute $N_b$ PBHMACHO binaries randomly
following the adopted HDP  in $r<D_{halo}$ and
assigning frequencies according to the PDF in equation (\ref{fnuapp})
for $\nu_{min}<\nu_{gw}<\nu_{max}$.

\item We make an observation in this numerically generated galactic
halo .
%We select the PBHMACHO binaries
%whose $SNR=h_i/h_{\nu} (\Delta \nu)^{1/2}$ is above a threshold,
%which is taken as $5$ here.
%Note that we cannot use all these binaries
Note that we cannot use all individual sources
to reconstruct the density profile,
since for low frequency  
a maximum distance to be observed as individual sources
in equation (\ref{dmax}) is short.
If we want to determine the HDP within $r<D_{obs}$,
we have to use binaries with frequencies $\nu_{gw}>\nu_{obs}$
where $\nu_{obs}$ is determined by
$D_{max}^{(2)}(\nu_{obs},0)=D_{obs}+D_0$.
Here $D_0=8.5$ kpc is
the distance from the galactic center to the earth.
%
%\item After taking into account of  observational errors,
%we can obtain the distribution map of the PBHMACHO binaries.
For simplicity,
we use a uniform probability distribution to
assign the observational error of $\epsilon_1+3\epsilon_2+\epsilon_3$
to the distance 
and the error of $3^{\circ}$ to the angular resolution.

\item By fitting the distribution map of the HDP,
we can compare the reconstructed HDP with the ``real'' HDP.
Note that observationally the normalization $n_s$ in equation (\ref{nr}) 
has to be replaced by the density of PBHMACHO binaries at the galactic center $n_{sb}$.
$n_s$ is obtained from  
$n_s=n_{sb}/F_b(\nu_{obs},\nu_{max})$.
\end{enumerate}

An example of simulated observations is shown in Figure 1.
This histogram shows the number $N_i$ of the observed PBHMACHO binaries
whose distances from the galactic center are in 
$(i-1) \delta r \le r < i \delta r$ $(i=1,2,\cdots)$.
We adopt $\delta r=1$ kpc
to determine the structure within a few kpc from the galactic center.
Here we set $T=10$ years and $D_{obs}=50$ kpc,
which corresponds to $\nu_{obs}=9.63\times 10^{-3}$ Hz.
In this realization,
$N_{map}$, the total number of PBHMACHO binaries
which can be used to determine the HDP,
is $719$.
In order to estimate the fitted parameters 
($n_{sb}$, $D_a$ and $\lambda$),
we apply the least squares method\footnote{
Strictly speaking, we may have to maximize 
the probability for observing $N_i$
PBHMACHO binaries in $i$-th bin from the Poisson distribution,
$P(n_{sb},D_a,\lambda)
=\prod_{i}\left\{
{{[n_i(n_{sb},D_a,\lambda)]^{N_i}}} e^{-n_i(n_{sb},D_a,\lambda)}
/{N_i !}
\right\},$ instead of minimizing $\chi^2$.
However, since almost all $N_i$ is larger than $10$,
it will be a reasonable assumption 
that the shape of the Poisson distributions
governing the fluctuations is nearly Gaussian.
}
minimizing $\chi^2 = \sum_i {{[N_i-n_i(n_{sb},D_a,\lambda)]^2}
/{\sigma_i^2}}$,
where
$n_i(n_{sb},D_a,\lambda)=\int_{(i-1)\delta r}^{i \delta r} 4 \pi n(r) dr$.
The variance $\sigma_i$ of $N_i$ can be estimated
by $\sigma_i = \sqrt{n_i(n_{sb},D_a,\lambda)}$,
since the distribution of $N_i$
will follow the Poisson distribution with mean
$n_i(n_{sb},D_a,\lambda)$
assuming that the statistical uncertainty dominates the
instrumental uncertainty due to the observational errors.
For this realization,
the fitted parameters turn out to be $n_s/n_s^{real}=0.779$,
$D_a=6.20$ kpc and $\lambda=1.04$,
where $n_s^{real}=2.60 \times 10^{-2}$ pc$^{-3}$ is the ``real'' value.
The reduced $\chi^2$ is $0.913$
with 47(=$D_{obs}/\delta r-3$) degrees of freedom.
In Figure 2,
the ``real'' parameter and the fitted parameter
are marked with a filled square and a cross respectively
in the $D_a$-$\lambda$ plane.
The contours of constant $\Delta \chi^2$ are also plotted
with $\Delta \chi^2=1.00,\ 2.30,\ 4.00$ and $6.17$.
In Figure 3, the ``real'' HDP and the reconstructed
HDP normalized by $n_s^{real}$
are shown.
It seems that in our method  HDP is reconstructed  quite well
except for the central region.

We performed $10^{4}$ simulations of observations
with (or without) the instrumental error to
obtain the probability distributions of the core radius $D_a$
and the power index $\lambda$.
The mean values $\langle w \rangle$ and 
the dispersions $\Delta w = (\langle w^2 \rangle -
\langle w \rangle^2)^{1/2}$ of these parameters $w$
with (or without) the instrumental error are shown in Table 1.
%We find that the instrumental error does not affect the results
From Table 1, we find that the instrumental error does not affect the results
so much.\footnote{
This also justifies the assumption that the statistical error dominates the
instrumental error.
Note also that the dispersions $\Delta w$ are consistent with 
the extent of the contour $\Delta \chi^2=1.00$ in Figure 2.
This justifies the assumption that the distribution of the number $N_i$
of the observed PBHMACHO binaries in $i$-th bin
is nearly Gaussian.}
The probabilities that these parameters $w$ are within
$|w-\langle w \rangle|<\Delta w$ and $2\Delta w$
turn out to be about 70\% and 95\% respectively from these realizations.
Although the power index $\lambda$ is determined within 10\% (20\%)
error in  89\% (99.7\%) confidence level (CL), respectively,  by ten years observation,
the dispersion of the core radius $D_a$
is somewhat large, 25\% (50\%) error in 63\% (93\%) CL.

After we know the power index $\lambda$
accurately by this global observation, 
we can analyze the HDP
for shorter distances $r<{\hat D_{obs}}<D_{obs}$
using the PBHMACHO bianries with lower frequencies $\nu_{gw}>\hat \nu_{obs}$,
where $\hat \nu_{obs}(<\nu_{obs})$ is determined by 
$D_{max}^{(2)}(\hat \nu_{obs},0)=\hat D_{obs}+D_0$
from equation (\ref{dmax}).
For example, for $T=10$ years and $\hat D_{obs}=10$ kpc, 
which corresponds to $\hat \nu_{obs}=5.15\times 10^{-3}$ Hz,
the mean value and the dispersion of
$D_a$ are found to be
$\langle D_a \rangle=4.81$ kpc and $\Delta D_a=0.710$ kpc
from $10^4$ realizations with $\delta r=0.5$ kpc
and $\lambda=1$.
The dispersion $\Delta D_a$ is reduced  by a factor 0.5.
Then, the core radius $D_a$ is determined within 25\% (50\%) error
in 91\% (99.8\%) CL.

\section{DISCUSSIONS}
In this paper we have quantitatively investigated
how well the HDP consisting of PBHMACHOs
can be determined by the observation of the low frequency GWs,
assuming the spherical HDP in equation (\ref{nr}).
We have found that ten years observation by LISA
will be able to determine  $\lambda$, the power index of the HDP,
within 10\% (20\%) error
and $D_a$, the core radius, within 25\% (50\%) error
in about 90\% (99\%) CL, respectively.

The halo of our galaxy may be non-spherical 
(e.g. Olling and Merrifield 1997). For a  non-spherical
halo if we calculate  quadrupole moments of positions of PBHMACHO binaries we can 
determine axial ratios ($\langle c/a \rangle$ and $\langle b/a \rangle$) 
of the dark halo (for
details see Dubinski \& Carlberg 1991).
Since about 700 PBHMACHO binaries can be used by ten years observation,
errors in the axial ratios are estimated as less than $10\%$
if the axial ratios are less than 0.8.

We have assumed that  MACHOs are PBHs.
However they may be  white dwarfs, or some other compact
objects (e.g. Freese et al. 1999). For such cases also
it may not be so strange  to expect that some of them are binaries.
If  a  fraction $\sim 10^{-6}$ of them 
is in binary systems emitting GWs
in the frequency range of $10^{-3}$ Hz $\siml \nu_{gw} \siml
10^{-1}$ Hz, similar arguments to this paper  will hold
even for non-black hole MACHOs (see also Bond \& Carr 1984).

\acknowledgments
We would like to thank professor H. Sato for continuous
encouragement and useful discussions. 
We are also grateful to K. Taniguchi, T. Harada and H. Iguchi
for useful discussions.
This work was supported in part by
Grant-in-Aid of Scientific Research of the Ministry of Education,
Culture, Science and Sports, No. 9627 (KI), No.09640351 (TN), 
No.09NP0801 (TN).

%
% Figure captions
%

\newpage 
\begin{figure}
\plotone{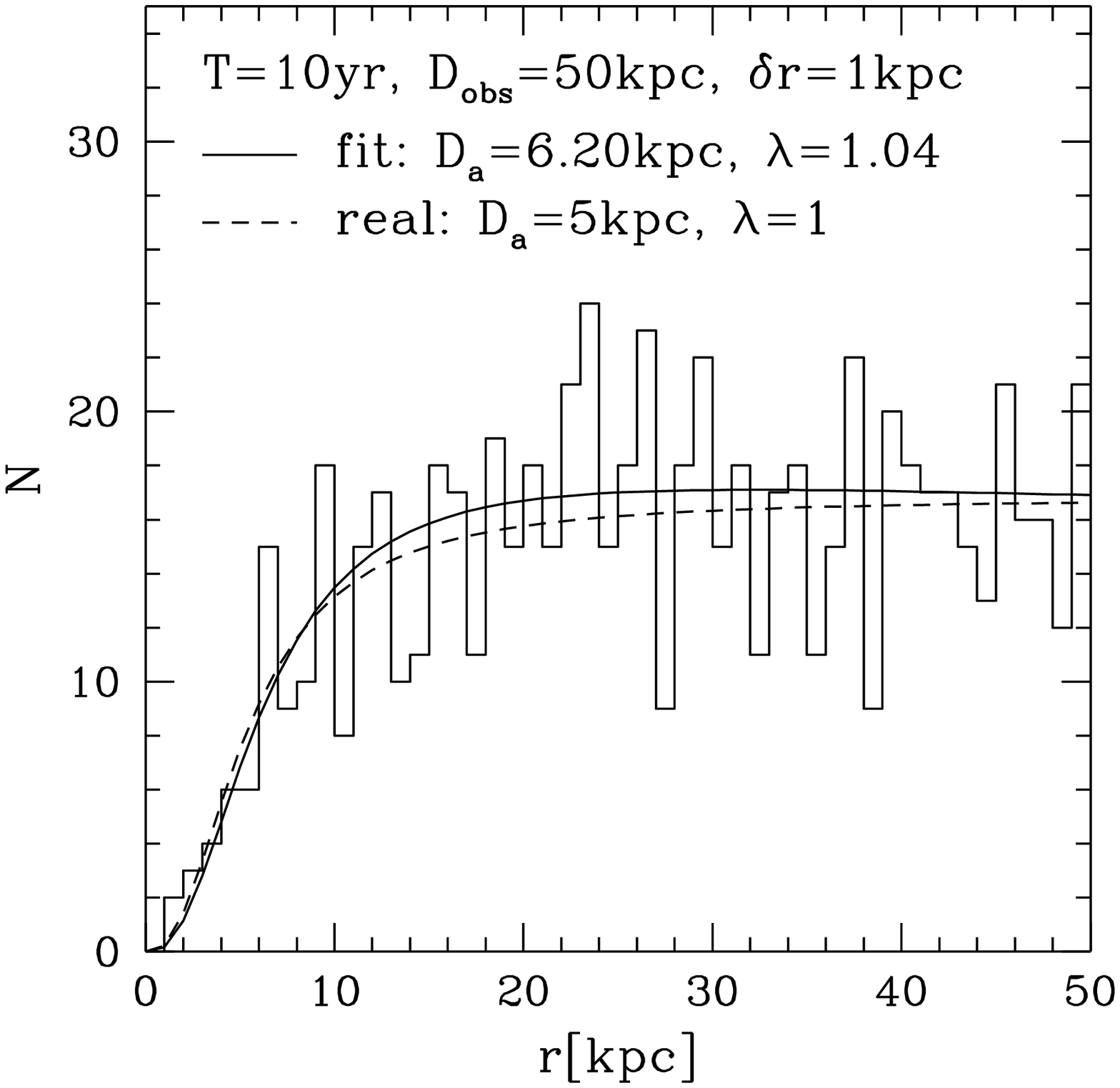}
\caption[fig1.ps]
%\figcaption
{The number $N_i$ of the observed PBHMACHO binaries
  whose distances from the galactic center are within 
  $(i-1) \delta r \le r < i \delta r$ $(i=1,2,\cdots)$ 
  in one experimental realization is shown.
  We adopt $\delta r=1$ kpc.
  Here we set $T=10$ years and
  $D_{obs}=50$ kpc,
  which corresponds to $\nu_{obs}=9.63\times 10^{-3}$ Hz.
  The fitted curve ({\it solid line}) 
  and the ``real'' curve ({\it dashed line}) are also shown.
  The fitted parameters are $n_s/n_s^{real}=0.779$,
  $D_a=6.20$ kpc and $\lambda=1.04$,
  where $n_s^{real}=2.60 \times 10^{-2}$ pc$^{-3}$ is the ``real'' value.
  The reduced $\chi^2$ is $0.913$
  with $47(=D_{obs}/\delta r-3)$ degrees of freedom.
  }
\end{figure}

\newpage 
\begin{figure}
\plotone{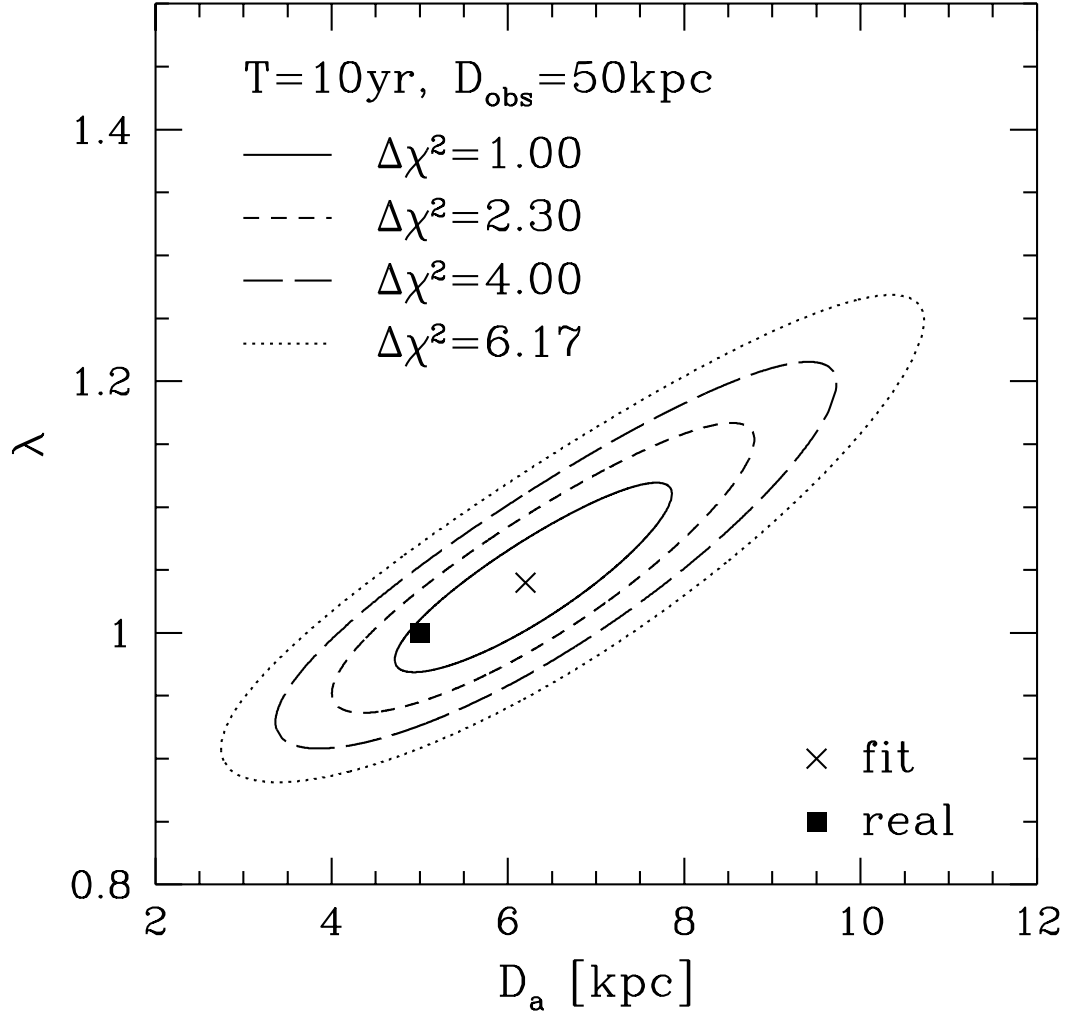}
\caption[fig2.ps]
%\figcaption
{The ``real'' parameter and the fitted parameter
obtained from one experimental realization in Figure 1
are marked with a filled square and a cross respectively
in the $D_a$-$\lambda$ plane.
The contours of constant $\Delta \chi^2$ are also plotted
with $\Delta \chi^2=1.00,\ 2.30,\ 4.00$ and $6.17$.
Note that for the Gaussian fluctuations
the projections of the contours $\Delta \chi^2=1.00$ and $4.00$
onto one axis contain $68.3\%$ and $95.4\%$
of data projected onto the axis respectively,
and the contours $\Delta \chi^2=2.30$ and $6.17$
contain $68.3\%$ and $95.4\%$ of data respectively.
}
\end{figure}

\newpage 
\begin{figure}
\plotone{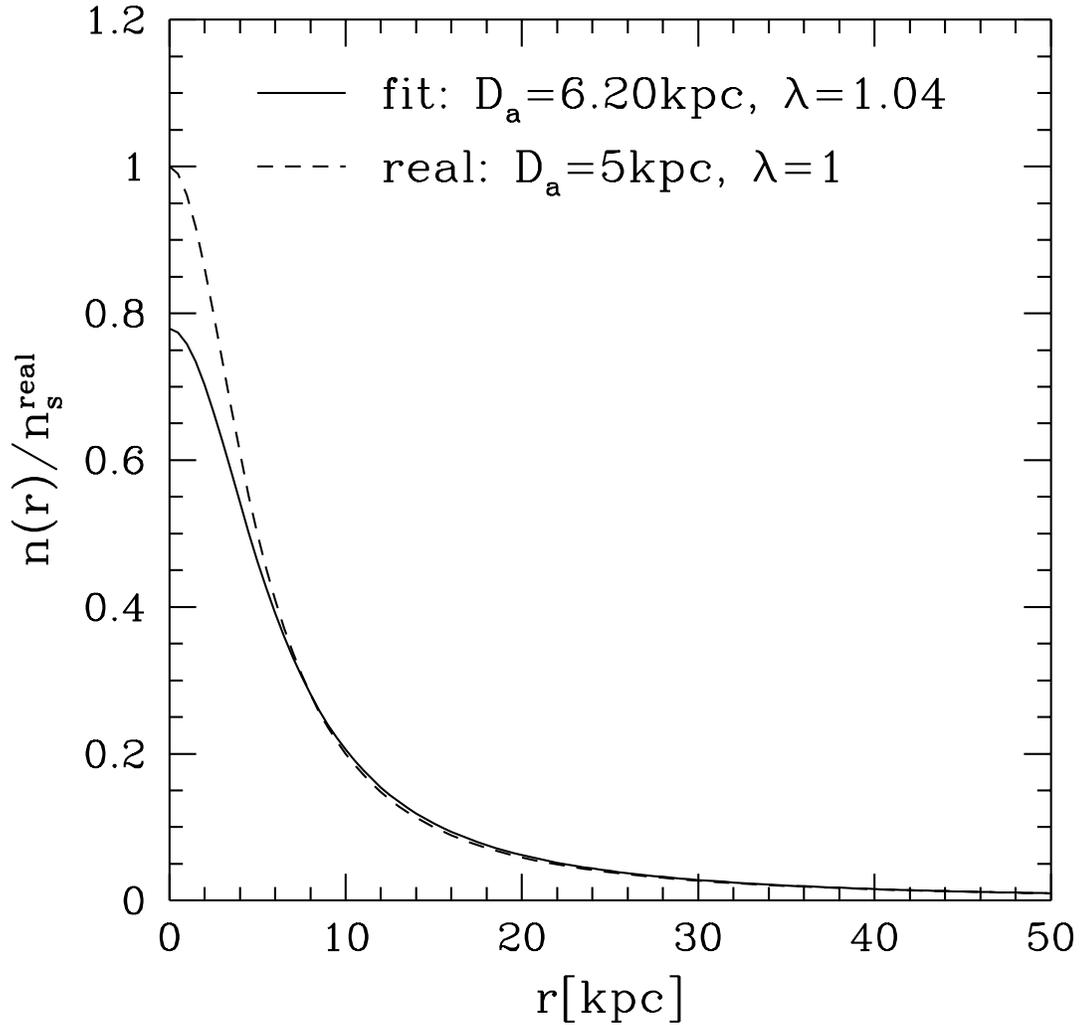}
\caption[fig3.ps]
%\figcaption
{The ``real'' density profile 
with $D_a=5$ kpc and $\lambda=1$ ({\it dashed line})
and the density profile 
fitted from Figure 1
with $n_s/n_s^{real}=0.779$, $D_a=6.20$ kpc and $\lambda=1.04$
({\it solid line}) are shown.
These density profiles are
normalized by $n_s^{real}=2.60 \times 10^{-2}$ pc$^{-3}$.
}
\end{figure}

%
% Tables
%
\newpage 
% Useful definitions

%\documentstyle[apjpt4]{article} % Comment to include the table in the manuscr.
%\pagestyle{empty}
% Override \tablevspace definition in apjpt.sty
\def\tablevspace#1{\noalign{\vskip#1}}

%\begin{document}                % Comment to include the table in the manuscr.
\begin{deluxetable}{lccccc}
%\footnotesize                  % Comment to get a camera-ready table
\tablewidth{0pt}
\tablenum{1}
\tablecaption{The mean values $\langle w \rangle$
  and the dispersions $\Delta w=(\langle w^2 \rangle -
  \langle w \rangle^2)^{1/2}$ of the core radius
  $D_a$ and the power index $\lambda$ obtained from
  $10^{4}$ experimental realizations
  with ({\it out of parentheses})
  and without ({\it in parentheses}) the instrumental error
  are shown for several observational time $T$.
  $\nu_{dis}$ is the minimum frequency of the binaries to which
  LISA can measure distances.
  $\nu_{obs}$ is the minimum frequency of the binaries which we can use
  to determine the density profile within $r<D_{obs}=50$ kpc.
  $\langle N_{map} \rangle$
  is the mean number of the PBHMACHO binaries that can be used to
  determine the density profile.
  }
\tablehead{
\colhead{$T$}                  & \colhead{$\nu_{dis}$}       &   
\colhead{$\nu_{obs}$}       &
\colhead{$\langle N_{map}\rangle$} &
\colhead{$\langle D_a\rangle \pm \Delta D_a$}  &
\colhead{$\langle\lambda\rangle \pm \Delta \lambda$}
\nl \tablevspace{5pt}
\colhead{[year]}                  & \colhead{[mHz]}   &
\colhead{[mHz]}            &
\colhead{} &
\colhead{[kpc]} &
\colhead{}
}
%\tablenotetext{a}{implied by $\delta_{\rm cf}$}
\startdata
2 & 4.08 & 15.1 & 217 (214) &
4.60 $\pm$ 2.44 (4.68 $\pm$ 2.43) & 1.00 $\pm$ 0.114 (1.01 $\pm$ 0.116) \nl
4 & 2.80 & 12.4 & 367 (363) &
4.57 $\pm$ 1.90 (4.67 $\pm$ 1.90) & 0.992 $\pm$ 0.0876 (1.00 $\pm$ 0.0894) \nl
6 & 2.24 & 11.1 & 496 (492) &
4.62 $\pm$ 1.64 (4.71 $\pm$ 1.64) & 0.992 $\pm$ 0.0763 (1.00 $\pm$ 0.0771) \nl
8 & 1.92 & 10.2 & 614 (608) &
4.65 $\pm$ 1.47 (4.74 $\pm$ 1.46) & 0.991 $\pm$ 0.0684 (0.999 $\pm$ 0.0689) \nl
10 & 1.70 & 9.63 & 723 (716) &
4.67 $\pm$ 1.35 (4.76 $\pm$ 1.34) & 0.990 $\pm$ 0.0626 (0.999 $\pm$ 0.0632) \nl
\enddata
\end{deluxetable}
%\end{document}                  % Comment to include the table in the manuscr.

\end{document}